\documentclass[sigconf]{acmart}

\usepackage{acronym}
\acrodef{SFINAE}{Substitution Failure Is Not An Error}

\copyrightyear{2020} 
\acmYear{2020} 
\setcopyright{acmlicensed}
\acmConference[CEP 2020]{Computing Education Practice 2020}{January 9, 2020}{Durham, United Kingdom}
\acmBooktitle{Computing Education Practice 2020 (CEP 2020), January 9, 2020, Durham, United Kingdom}
\acmPrice{15.00}
\acmDOI{10.1145/3372356.3372357}
\acmISBN{978-1-4503-7729-4/20/01}

\begin{document}

\title{Computing with CodeRunner at Coventry University}
\subtitle{Automated summative assessment of Python and C++ code}

\author{David Croft}
\affiliation{%
  \institution{Coventry University}
  \streetaddress{}
  \city{Coventry}
  \state{U.K.}
}
\email{David.Croft@coventry.ac.uk}

\author{Matthew England}
\affiliation{%
  \institution{Coventry University}
  \streetaddress{}
  \city{Coventry}
  \state{U.K.}
}
\email{Matthew.England@coventry.ac.uk}

\renewcommand{\shortauthors}{D. Croft and M. England}

\begin{abstract}
CodeRunner is a free open-source Moodle plugin for automatically marking student code.  We describe our experience using CodeRunner for summative assessment in our first year undergraduate programming curriculum at Coventry University.  We use it to assess both Python3 and C++14 code (CodeRunner supports other languages also).  
We give examples of our questions and report on how key metrics have changed following its use at Coventry.  
\end{abstract}



\maketitle

\section{Introduction}

CodeRunner is a tool developed at the University of Canterbury, New Zealand, for automatically assessing student code \cite{LH16}.  We use it for summative assessment of programming of first year Computer Science students at Coventry University.  

The use of automated assessment for coding is of course not a new topic $-$ see for example the survey \cite{AlaMutka2005}.  However, it seems historically there are a wide range of independent tools with few used very widely beyond the institutions which created them.  In contrast, CodeRunner was developed as a plugin for the very widely used learning management system Moodle\footnote{According to the following 2017 report Moodle is used by an absolute majority of HE institutions in Europe, Latin America and Oceania; and a quarter in North America: \url{http://eliterate.us/academic-lms-market-share-view-across-four-global-regions/}}.  Further, it is fully approved by Moodle which means installation is easy, both technically and in terms of administration\footnote{The approved plugin status meant our IT team team would add it to our Moodle installation without any major review.}.  CodeRunner has the potential to become the standard automated code assessment tool of choice, but to the best of our knowledge the only publication on CodeRunner is the 2016 introduction by the developers in ACM Inroads \cite{LH16}.  We hence think there is value in sharing our experiences.

\section{Our Context and Motivation}

The authors design and lead the first year programming curriculum for the Computer Science (CS) degree at Coventry University.  This is based around two key modules\footnote{These modules ran for the first time in 2018/19.  Later references to what happened in prior years refer to two very similar modules which these replaced.}:
\begin{description}
\item[4000CEM Programming and Algorithms] First semester.  \\Introduces basic control structures in Python 3 and ideas around different algorithms to solve the same problem.
\item[4003CEM Object Oriented Programming] Second semester.  Translates from Python to C++14; then introduces the ideas of objects, classes, and inheritance.
\end{description}
Our intake is diverse: we do not require A-level Mathematics, prior CS study, or programming experience (although many students will have some or all of these).  The 2018/19 cohort had $\sim$280 students.

\subsection{Additional Project Modules}

These modules focus on core programming ideas and theory, demonstrated by fairly traditional programming exercises and tasks (albeit delivered in a modern cloud-based environment as described below).  However, in each semester students also take an additional group project module where they apply all the concepts they are learning in their degree to a real world problem via an activity-led learning approach.  These projects, and their unique administration and assessment, are described in \cite{BE20}.  

\subsection{Other Recent Innovations at Coventry}

The authors described in CEP 2019 how they have introduced the learning environment Codio to their modules \cite{CE19}.  Codio provides students with online virtual Linux boxes, which staff equip with guides.  In \cite{CE19} we described how the adoption of Codio was, in main, a response to a low-level of formative feedback provision and uptake (being made harder still by rapidly increasing student numbers).  Our Codio units contain numerous formative tasks that feedback on student code automatically by running test scripts written by the authors.   Additional benefits from Codio described in \cite{CE19} include a standardised development environment on diverse student hardware, cloud storage for student code, and a detailed source of additional student data on engagement and progress. 
 
\subsection{Assessment Prior to CodeRunner}

As part of the project modules, students give group reports, presentations, and individual vivas on their programming.  Modules 4000CEM and 4003CEM are then assessed using tests: each has one in the middle of the term and one at the end. It is the mid-term tests where we have made a change to employ CodeRunner.  

Given the mid-semester time, tight feedback deadlines, and large quantity of students, an auto-marked mid-term was essential (especially with the human feedback committed to the projects).  We use Moodle quizzes, but prior to 2018/19 these employed only the standard Moodle question types like multiple choice.  
Such questions can assess knowledge of programming concepts and theory, but not the ability to formulate an algorithm and code to solve a problem.  4000CEM has also a written exam which does assess these abilities, however, this is an inauthentic environment; different to both the one in which students learnt the material and any professional environment in which they would later apply it. E.g. no syntax checker, no access to language documentation, no ability to edit the code without writing it all out again.  The additional verbosity required for C++ code meant that 4003CEM has no written exam.
Our positive results with automated formative feedback \cite{CE19} led us to consider the same approach for summative assessment.

\subsection{Codio for Summative Assessment?}

Standard Codio tasks are administered by code stored on the VM to which students have sudo rights: there is motivation to edit this for formative assessment, but it would be an unacceptable risk for summative. Further, our university regulations forbid the use of a third-party in summative assessment, which also precluded the use of any alternative providers with more locked down environments.  We thus looked for another method of automated code testing for summative assessment and found CodeRunner. 
Codio recently implemented secure assessments, which are not stored on the student VM, but our third-party restriction still applies. 

\section{CodeRunner}

CodeRunner is a free tool for testing student code\footnote{\url{http://coderunner.org.nz}}.  It is an approved plug-in for the widely used learning management system Moodle and is also available open source for those who wish to customise\footnote{\url{https://github.com/trampgeek/moodle-qtype_coderunner}}.

We use CodeRunner with Python3 and C++14, but the standard installation also has question types for C, Java, Javascript (NodeJS) PhP, Matlab, and Octave; and the system essentially allows for use with any language that can be executed on the Linux server.

\subsection{Main Functionality}

CodeRunner provides an additional Moodle quiz question type\footnote{Thus CodeRunner questions can be combined in a test with those of other types.} where the student answer is code, sent to a separate (university) server where it is executed against unit tests.  The tests can simply check for desired standard output, or perform something more involved which the teacher codes up.  Some of our favourite features:
\begin{itemize}
\item Students type code into a box with basic IDE functionality, e.g. syntax highlighting and automated indentation.
\item Students can use a syntax checker before the full unit tests.
\item If student code creates error messages they are shown.
\item After seeing the results of the unit tests students can edit their code and try again.  The number of attempts, and penalty regime (if any) is fully customisable. 
\item Students can be given links to resources such as the standard documentation for a language.
\end{itemize}
Thus a CodeRunner test an \emph{authentic} examination of programming ability, and we can still maintain test integrity.  Students take tests under exam conditions with the usual Moodle quiz security tools:  passwords, required IP, use of the Safe Exam Browser, etc.

\subsection{Options for Individual Unit Tests}

The number of marks for each unit test can be varied.  Individual tests can also be flagged as one of the following special cases:
\begin{itemize}
\item \textbf{Examples:} meaning the expected input and output is displayed prior to the student attempt.  These can aid student's understanding of what is being asked of them.
\item \textbf{Pre-check:} meaning these tests are run for free (i.e. without penalty).  We use these to check the student has spelt the function name correctly, and to inform them of syntax errors.
\item \textbf{Hidden:} meaning students do not see the input and output even after the tests.  Essential to avoid a student hard-coding a large if-statement with all the known test inputs!
\end{itemize}

\subsection{Other Useful Features}

There is an option to pre-populate the answer box with partially complete / buggy code which you can then ask the student to extend / fix.  This is particularly useful for assessing object oriented programming: e.g. asking students to implement an additional method which interacts with the other methods and attributes in specified ways.  Students can reset the answer box to its original state.

A teacher can opt to store their own model answer to a question.  This can then be revealed to students after the test.  It can also be evaluated against the unit tests after each change $-$ a very useful error check during the development of quizzes and their tests.

\section{Examples}

\subsection{Python}

\begin{figure}
\includegraphics[width=8.5cm]{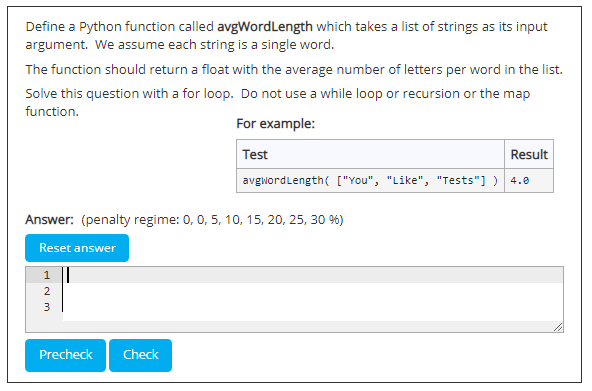}
\caption{Example of a CodeRunner question \label{fig:Py1}}
\Description[Screenshot of an example CodeRunner question.  ]{Screenshot of an example CodeRunner question with the box underneath for students to enter their code.  The questions reads as follows:  Define a Python function called avgWordLength which takes a list of strings as its input argument.  We assume each string is a single word.  The function should return a float with the average number of letters per word in the list.  Solve this question with a for loop.  Do not use a while loop or recursion or the map function.}
\end{figure}

\begin{figure}
\includegraphics[width=8.5cm]{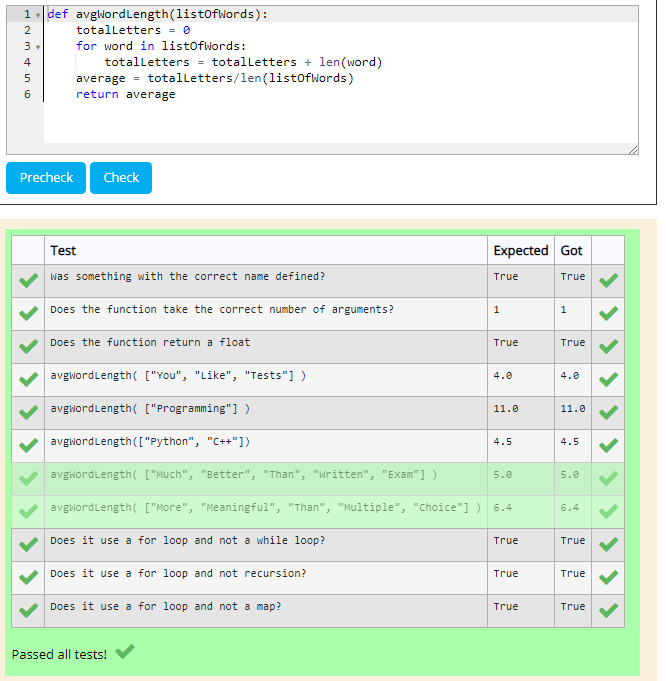}
\caption{Example of a student answer to the question in Figure \ref{fig:Py1} that would pass all of tests (the opaque rows are hidden tests which would not be seen by the student)\label{fig:Py2}}
\Description[Screenshot of code which solves the example question]{Screenshot of code which solves the example question.  The code defines the function, initialises a counter at 0, runs a for loop over the list, in each iteration increases the counter by that number of words, then after the loop divides the counter by the number of words and returns this as output.}
\end{figure}

Figure \ref{fig:Py1} shows a typical question from the assessment of our Python module.  Students must write a function to calculate the average string length for a given list of strings by iterating with a for loop.  The expected correct answer would be similar to the one shown passing all tests in Figure \ref{fig:Py2}.
Our tests mark this question out of 10 marks as follows (in the order of Figure \ref{fig:Py2}):
\begin{description}
\item[0.5] for defining a function with the correct name.  Checked with the Python code \texttt{"avgWordLength" in globals()}
\item[0.5] for the function have the correct number of parameters.  Checked using the \texttt{signature} function of the \texttt{inspect} module of the Python Standard Library.
\item[0.5] for the function returning data of the correct type (float).
\item[0.5 $\times$ 3] for three test cases for which the student can observe the input and output.  The first of these is flagged as an \textit{Example} and so was included with the question in Figure \ref{fig:Py1}.
\item[2.0 $\times$ 2] for two further test cases which are flagged as hidden (they appear opaque when a teacher views as in Figure \ref{fig:Py2}).
\item[1.0 $\times$ 3] for checks that the requested control structure was used.  The student code is stored as a string so we just check for substrings \texttt{" for "}, \texttt{" while "} \& \texttt{" map "}\footnote{The spaces around the words are deliberate as we care only for these Python keywords and their appearance within longer words.  Python's required indentation for function body means they cannot occur at the start of a line.} and check recursion by counting occurrences of the function name.
\end{description}

\begin{figure*}
\includegraphics[width=\textwidth]{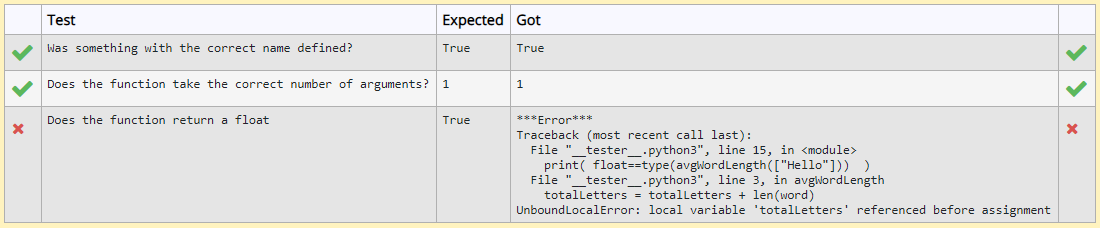}
\caption{Test results for a student answer that caused a runtime error \label{fig:Py3}}
\Description[Screenshot of test results for student answer causing runtime error]{Screenshot of test results for student answer causing runtime error.  Testing stops after one triggers the error.}
\end{figure*}

\begin{figure}
\includegraphics[width=8.5cm]{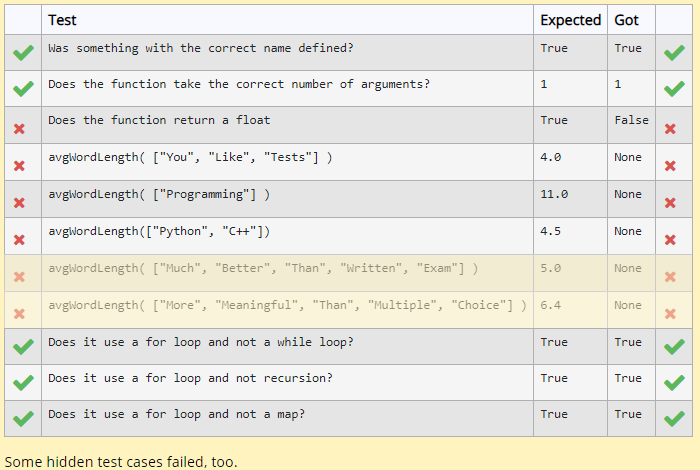}
\caption{Failed unit tests without raising errors  \label{fig:Py4}}
\Description[Screenshot of test results for a student answer that is incorrect but triggers no error messages]{Screenshot of test results for a student answer that is incorrect but triggers no error messages. All unit tests are run.}
\end{figure}

\noindent An initial attempt at a solution to the problem may be the following:\\
\includegraphics[trim={1.8cm 14.3cm 11cm 12.1cm},clip]{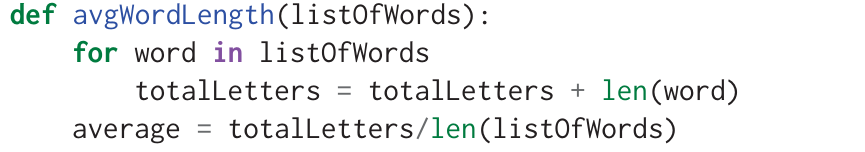}
\\
\noindent This code contains 3 errors:  (1) missing colon at the end of line 2; 
(2) no initialisation of the variable \texttt{totalLetters}; (3) not returning the final answer. 
(1) is a syntax error:  the Precheck button would display the Python interpreter message pointing to the end of line 2.  
Once we add the colon the PreCheck would return no further information $-$ students should then use Check to get the test results in Figure \ref{fig:Py3}. The testing stops after one triggers a runtime error with students shown the usual message.  
Initialising \texttt{totalLetters} and checking again gives the results in Figure \ref{fig:Py4}.  This time there are no runtime errors so all unit tests are evaluated.  Students do not see the opaque hidden test details but are informed of their failure.  

If students then added in the return statement and pressed Check again they would pass all tests as in Figure \ref{fig:Py2}.  They would have only used Check twice and thus receive no penalty, getting full marks for the question.  This reflects our view that errors (1)-(3) are not critical gaps in student knowledge, \textit{so long as they are able to read the error messages and unit test feedback to correct them quickly}.

\subsubsection*{Gaming the system?}

A function that simply returned 4.0 (the example result) would get 20\%, while one with an if-statement for the three visible test answers with any float otherwise could obtain 30\%.  Similarly, students could randomly guess the right answer on a multiple choice question.
We could follow that if-statement with a syntactically correct but meaningless for-loop to pass all but the hidden tests and score 60\%.  However, in that case the student has actually demonstrated knowledge of Python for-loops: it is unlikely they would posses this knowledge but still need to game the system.

\subsection{C++}

Utilising CodeRunner in the assessment of C++ code poses additional challenges when compared to Python. It is likely that these issues will also manifest in the assessment of other statically typed and compiled languages (e.g. Java).  
Essentially, a range of student mistakes which would cause runtime errors in Python cause compiler errors in C++.  Then, as compilation is unsuccessful, it is impossible to run the unit tests and so students receive no marks for what may have been correct code with just minor errors.

In the code below the student was meant to declare a function that doubled its argument but misunderstood and wrote a function to simple return 10 (probably this was the example case). 
\\
\includegraphics[trim={11.0cm 6.8cm 1.8cm 18.7cm},clip]{P3}
\\
\noindent Of course, the student does not deserve many marks but they have shown they can create a C++ function and so if marked by hand it would get some partial credit.  However, a unit test like that shown on the final line would cause a compilation error.  
It is a trivial example but the issue applies generally: e.g. if a student declared a class with correct inheritance properties then awarding zero on the basis on a minor error in one method is extremely undesirable.


To address this we utilise a combination of \ac{SFINAE} \cite[\S~8.3.1]{VJ02} and preprocessor macros to perform code introspection during compilation.  A simplified demonstration of \ac{SFINAE} is shown in the code below where we define a series of increasingly more specific templates to test the existence of the required function with a signature increasingly closer to the correct one.  This approach allows for partial credit, and also more intuitive error messages than a C++ complier.   \\
%
\includegraphics[trim={1.8cm 6.2cm 11cm 17.1cm},clip]{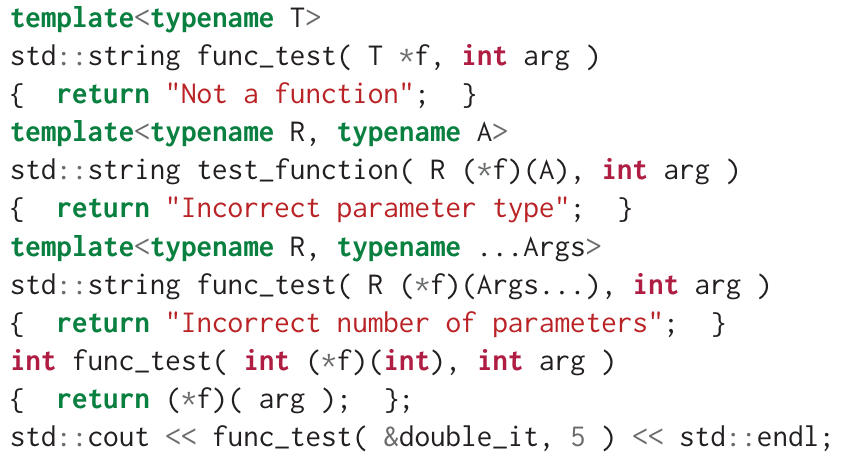}

We have developed C++ introspection code to determine: the existence of functions, parameter number/types, return types, class inheritance/attributes/methods, and more.  It was a significant time investment but mostly a one-off payment\footnote{Contact the first author for access to the C++ Library of introspection code.}.
Student code may still cause error messages: but these errors are from within the student's code (rather than the test code) and so the error message should be enough to correct them.


\section{Results at Coventry}

In both modules our CodeRunner assessments had lower average marks and pass rates that the multiple choice tests they replaced.  In our opinion this is because the multiple choice tests overrated student's programming ability.  The final written exam for 4000CEM has been of a similar format and difficulty for several years - the cohort who were assessed in the mid term using CodeRunner actually had average exam mark 7\% higher and pass rate 10\% higher!

We measure student satisfaction with an online questionnaire that mimics the UK National Student Survey.  In 4000CEM overall satisfaction increased 4\% to 93\% following the introduction of CodeRunner for the mid term assessment.  More importantly, students satisfaction with assessment and feedback increased 7\% to 90\%.  Assessment and feedback is often the area where students are most dissatisfied so  this result is particularly promising.  More generally, the literature suggests that the \emph{relative correctness} of work implies by unit tests is accepted by the student body \cite{Kolikant2005}.

In 4003CEM the fall in initial student marks was more pronounced and there was a corresponding fall in student satisfaction.  This prompted the development of the introspection code to better reward partial credit described above\footnote{Approved and applied retrospectively by the exam board for 2018/19.}.  With this in place we anticipate a smoother experience assessing C++ next year.  One topic for our future work with CodeRunner is an investigation on the most appropriate penalty regime to apply for checking the unit tests.

\begin{acks}
We are grateful to the developers of CodeRunner; and the organisers of the 2017 CodeRunner workshop at the University of Edinburgh.
\end{acks}

\bibliographystyle{ACM-Reference-Format}
\bibliography{EdRes}

%
%

\end{document}